\documentclass[final,1p,times]{elsarticle}

\usepackage{amsmath}
\usepackage{amsfonts}
\usepackage{color}
\usepackage{graphicx}
\usepackage{makeidx} 
\usepackage[T1]{fontenc}
\usepackage{url}
\usepackage{float}

\usepackage[]{mcode}

\usepackage{textcomp}

\newcommand{\sgn}{\operatorname{sgn}}
\newtheorem{algorithm}{Algorithm}

\begin{document}

\begin{frontmatter}

\title{ Numerical proof of existence of fractional powers of Wiener processes }

\author[mbda]{Marco~Frasca\corref{cor1}}
\ead{marco.frasca@mbda.it}

\author[selex]{Alfonso~Farina}
\ead{alfonso.farina@selex-es.com}

\address[mbda]{MBDA Italia S.p.A.\\ 
Divisione Seeker\\
Via Carciano, 4-50, 60-70\\ 
00131 Roma (Italy)}

\address[selex]{Selex ES S.p.A.\\
Via Tiburtina km. 12,400\\
00131 Roma (Italy)}

\cortext[cor1]{Corresponding author.}
%


\date{\today}

\begin{abstract}
Using the Euler--Maruyama technique, we show that a class of Wiener processes exist that are obtained by computing an arbitrary positive power of them. This can be accomplished with a proper set of definitions that makes meaningful the realization at discrete times of these processes and make them computable. Standard results from It\=o calculus for integer powers hold as we are just extending them. We provide the results from a Monte Carlo simulation with a large number of samples. We yield evidence for the existence of these processes by recovering from them the standard Brownian motion we started with after power elevation. The perfect coincidence of the numerical results we obtained is a clear evidence of existence of these processes. This could pave the way to a generalization of the concepts of stochastic integral and relative process.
\end{abstract}





\end{frontmatter}


\section{Introduction}
\label{sec0}

Recently, we introduced a class of stochastic processes that could help to derive the Schr\"odinger equation as it also happens to the heat equation for the Brownian motion \cite{far2,fra1}. The existence of these processes has been questioned by some mathematicians claiming that the corresponding sums are not $L^2$--summable: It is always possible to find a discretization on the time interval such that the limit of the even smaller intervals makes Riemann sums diverging. A way out to this problem was devised in \cite{fra1}. Indeed, treating diverging sums as always happens in physics, i.e. attributing a finite value to them as yielded in \cite{hard}, can be satisfactory from a physicist standpoint but surely will leave mathematicians with the surviving firm belief that the objection still stands.

Stochastic processes have the peculiarity to be very easy to generate on computers and simulations can be realized making this the natural laboratory to test any relevant hypothesis on them. So, if one wants to prove the existence of a stochastic process, the best way to show it is with numerical computations. Indeed, there is a wealthy number of numerical codes available as exemplified in \cite{high}. Numerical algorithms, giving finite and meaningful results, become also the corresponding mathematical definition for the processes we aim to study.

The strategy of this paper is to start simulating a standard Brownian motion or Wiener process by integrating the corresponding stochastic differential equation (SDE) with a numerical technique (we have chosen the Euler--Maruyama method \cite{high}). Then, we define the $\alpha$--root of this process through the solution of the corresponding SDE with the Euler--Maruyama method. We recover the original process by elevating to the power $\alpha$ the random jumps obtained by this integration. These jumps must coincide with the jumps of the Brownian motion we started with. If this happens, the corresponding $\alpha$-- root process of the considered Brownian motion is proved to exist. We will see that this is indeed the case. For the sake of simplicity we will limit the study to the case $\alpha=1/2$, the square root process. 

In ref.~\cite{far2}, an equation for the square root process was provided. We will check this equation considering it as a SDE and we will solve it with the Euler--Maruyama method. An explicit definition of all the quantities that enter in this formula will be provided. We will see that this equation is able to recover exactly the original Brownian motion when squared 
with a sign process properly removed.
When it is seen as a properly defined identity for the starting Brownian motion, this equation performs excellently well reproducing the original Brownian motion all the times. 
So, this formula grants an excellent representation for the square root process. But we will note how one can dispose of the sign process that alters the scale of the original Brownian motion. As our aim in this paper is just to prove the existence of fractional Wiener processes by a Monte Carlo simulation, we will not deepen this matter here.

As we will see repeatedly in this paper, the theoretical analysis presented in \cite{far2} has the problem that the square root of a Wiener process, when squared to get back the original Brownian motion, leaves a sign process multiplied by an arbitrary constant altering the scale of the original process. Here we present a solution to this problem using the spin matrices and we will obtain a Wiener process belonging to a Clifford algebra easily generalizable to more complex algebras if needed.  

The paper is organized as follows. In Sec.~\ref{sec2} we discuss all the theoretical aspects of our study and introduce the proper definitions with the strategies we will adopt in the proof. In Sec.~\ref{sec3} we present the numerical results with the computer code to use to obtain them. In Sec.~\ref{sec4} we show how to get rid of a sign process in the square root of a Wiener process when squared using spin matrices. Finally, in Sec.~\ref{sec5} we yield the conclusions.

\section{Theoretical analysis}
\label{sec2}

Consider a class of stochastic equations that, in their simplest form, are
\begin{equation}
\label{eq:palf}
    dX = (dW)^\alpha.
\end{equation}
with $\alpha>0$ and $W$ a Wiener process \cite{okse}. These are defined through the Euler--Maruyama method in the discrete form
\begin{equation}
\label{eq:srnum}
    X_i=X_{i-1}+(W_i-W_{i-1})^\alpha
\end{equation}
with $i=0\ldots N-1$ such that the time interval $t\in [0,T]$ of the stochastic process is broken in $N$ parts with increments $T/N$. Also from It\=o calculus \cite{okse} one has: $(dW)^2=dt$, $dW\cdot dt=0$, $(dt)^2=0$ and $(dW)^\alpha=0$ for $\alpha>2$. Our aim is to verify that these processes exist as also the particular formula proposed in \cite{far2,fra1}
\begin{equation}
\label{eq:sroot}
    dX=(dW)^\frac{1}{2}=\left(\mu_0+\frac{1}{2\mu_0}|dW|-\frac{1}{8\mu_0^3}dt\right)\cdot\Phi_\frac{1}{2},
\end{equation}
being $\mu_0\ne 0$ an arbitrary scale factor and $\Phi_\frac{1}{2}=\frac{1-i}{2}\sgn(dW)+\frac{1+i}{2}$ a Bernoulli process equivalent to a coin tossing and the process $\sgn(dW)$ will be defined below for its operational implementation as the process $|dW|$. This process can be quite different for $\alpha\ne 1/2$. The arbitrary scale factor plays a crucial role in recovering the standard Brownian motion we started with. It is
\begin{equation}
\label{eq:Xsq}
    (dX)^2=\mu_0^2\sgn(dW)+dW
\end{equation}
with the definitions we will give below and in the source code of the computer simulation. The sign process multiplied by the scale factor $\mu_0^2$ must be removed to exactly recover the original Wiener process. One can get rid of it by using a Clifford algebra (we show this in Sec~\ref{sec4}). We note that we can define the equation on the rhs of eq.~(\ref{eq:sroot}) through the Euler-Maruyama method at discrete times yielding a clear definition of this formula as we are going to see in a moment.

We note that, for $\alpha=1$, Euler-Maruyama method reduces just to a cumulative sum that is, $X_0=W_0,X_1=W_1,X_2=W_2,\ldots$ and this is a standard way to simulate a standard Brownian motion: The technique is to generate the jumps $X_i-X_{i-1}$ by a normal distribution with $\sigma=\sqrt{\Delta t}$ and then to compute the cumulative sum as said. In this way we get immediately a numerical definition for the most general case (\ref{eq:palf}) as follows
\begin{equation}
   X_0=W_0^\alpha,\quad X_1=W_0^\alpha+(W_1-W_0)^\alpha, \quad X_2=W_0^\alpha+(W_1-W_0)^\alpha+(W_2-W_1)^\alpha,\quad \ldots.
\end{equation}
So, to give a meaning to eq.(\ref{eq:sroot}) we can introduce the following definitions through the Euler--Maruyama method for the given Brownian motion to be implemented in the computer simulation:
\begin{equation}
   \sgn(dW)=\{\sgn(W_0),\sgn(W_1),\sgn(W_2),\ldots\}.
\end{equation}
On a finite time $T$, for $N$ instants, this will be a string of $\pm 1$ values and it is a random variable with a Bernoulli distribution having $p=1/2$. Similarly,
\begin{equation}
\label{eq:phi}
   \Phi_\frac{1}{2}=\frac{1-i}{2}\{\sgn(W_0),\sgn(W_1),\sgn(W_2),\ldots\}+\frac{1+i}{2}
\end{equation}
on the same finite time $T$ at discrete instants. Finally, one has
\begin{equation}
    |dW| = \{|W_0|,|W_1|,|W_2|,\ldots\}
\end{equation}
for a finite time $T$ at discrete instants. We just note that one must have
\begin{equation}
    |dW|\sgn(dW)=\{|W_0|\cdot\sgn(W_0),|W_1|\cdot\sgn(W_1),|W_2|\cdot\sgn(W_2),\ldots\}=dW.
\end{equation}
In this way our code is able to recover the original Brownian motion making the squared power numerically, provided the aforementioned scale factor $\mu_0$ is taken into account. 
There is another way to see eq.~(\ref{eq:sroot}) and it is to interpret it as a SDE and solve it with the Euler--Maruyama method. We write
\begin{equation}
\label{eq:emsroot}
     X_{i}=X_{i-1}+\left(\mu_0+\frac{1}{2\mu_0}|W_i-W_{i-1}|-\frac{1}{8\mu_0^3}dt\right)\cdot\tilde\Phi_{\frac{1}{2},i}.
\end{equation}
This equation can be immediately inserted into a code for a computer simulation. This is what we have done. We will discuss it in the next section but we can show from this how to recover the standard Brownian process. Firstly we write
\begin{equation}
    \tilde\Phi_{\frac{1}{2},k}=\frac{1-i}{2}\{\sgn(W_k-W_{k-1})\}+\frac{1+i}{2}
\end{equation}
with $k$ running from $0$ to $N-1$. Note that this is different from the definition (\ref{eq:phi}). Now, we take the square of the jump $X_i-X_{i-1}$ to obtain
\begin{eqnarray}
    (X_i-X_{i-1})^2&=&\mu_0^2\sgn(W_i-W_{i-1})+\frac{1}{4\mu_0^2}(W_i-W_{i-1})^2\sgn(W_i-W_{i-1}) \nonumber \\
		&&-\frac{1}{4\mu_0^2}dt\sgn(W_i-W_{i-1}) \nonumber \\
		&&+\frac{1}{64\mu_0^4}(dt)^2\sgn(W_i-W_{i-1}) \nonumber \\
		&&+(W_i-W_{i-1})-\frac{1}{8\mu_0^4}dt(W_i-W_{i-1}).
\end{eqnarray}
The only relevant contributions come from the term $\mu_0^2\sgn(W_i-W_{i-1})+(W_i-W_{i-1})$ the others being negligibly small in agreement with expectations from It\=o calculus (we will see this in the next section with a Monte Carlo simulation). From eq.(\ref{eq:sroot}) we see that this is enforced by the condition $|\mu_0|>1/2$ as this grants that the coefficients become increasingly small. So, eq.(\ref{eq:emsroot}) yields an excellent representation of the square root of a standard Brownian motion but a contribution coming from the constant $\mu_0$ survives. 

In order to recover the original path from the given solution of the SDE (\ref{eq:srnum}) we work in the following way:
\begin{algorithm}
\label{alg:BBP}
Back to the Brownian path:
\begin{table}[H]
  \centering
  \begin{tabular}{l}
	\hline
   Compute the jump for the integrated Brownian motion: $J_i=W_i-W_{i-1}$. \\
	\\
   Compute the jump for the $\alpha$--process: $S_i=(X_i-X_{i-1})^\frac{1}{\alpha}$. \\
	\\
   Multiply $S_i$ by the sign of $J_i$ if needed. \\
	\\
   It must be $S_i=J_i$. \\
	\\
   Final step: Take the cumulative sums of $\{S_i\}$ or $\{J_i\}$ to recover the Brownian path $\{W_i\}$.\\
  \hline
  \end{tabular}
\end{table}
\end{algorithm}
The implementation of this algorithm will be seen in the next section.

 
\section{Numerical results}
\label{sec3}

We have implemented all the definitions introduced earlier and the algorithm \ref{alg:BBP} with a Monte Carlo simulations of 10000 independent Brownian paths. We simulated each single Brownian path with the \texttt{randn} function in MATLAB\textsuperscript\textregistered. We used these paths to evaluate eq.~(\ref{eq:sroot}) for each of them. We solved the SDEs, both for the Brownian motion and its square root, with the Euler--Maruyama method as shown in \cite{high}. We compared the jumps from the Brownian motion and its square root from the numerical solutions as in algorithm \ref{alg:BBP}. The MATLAB\textsuperscript\textregistered\ code is the following.

\begin{lstlisting}[frame=lines, caption=sdeMC.m]
%% Monte Carlo simulation of the square root extraction of a Brownian motion (BM)
%
%  sdeMC.m
%
%  We use the Euler-Maruyama integration method of SDE to define the square
%  root (SR) of a given Brownian motion. In order to check the correctness 
%  of the results, we perform a Monte Carlo simulation with 10000 independent 
%  paths.
%
%  Bibliography:
%
%  A. Farina, M. Frasca, M. Sedehi, 
%  ``Solving Schrödinger equation via Tartaglia--Pascal triangle: 
%   a possible link between stochastic processing and quantum mechanics'', 
%   Signal, Image and Video Processing, Volume 8, Number 1, 27--37, (2014).
%
%  M. Frasca, 
%  ``Quantum mechanics is the square root of a stochastic process'', 
%   arXiv:1201.5091 [math-ph].
%
%
%% Initialization
clear all
close all

w0 = 0;        % Initial condition

N = 2^8;       % Number of steps
M = 10000;     % Number of paths

dt = 1/N;      % Time step

t=0:dt:1;      % Time [0,1]

mu0 = 30;      % Scale factor in the square root of the Brownian motion

%% Body of the simulation

% Loop on the number of paths
for k=1:M

% Set initial values
w(k,1) = w0;
z(k,1) = w0;
v(k,1) = w0;

% Gaussian jumps
dW = sqrt(dt)*randn(1,N);
% Brownian motion
W(k,:) = cumsum(dW);
% Signs of the Brownian motion
dS1(k,:) = sign(W(k,:));                    % Frasca-Farina-Sedehi formula
dS(k,:) = sign(dW);                         % Frasca-Farina-Sedehi (FFS) SDE
% Bernoulli process
dF1(k,:)=(1-sqrt(-1))*dS1(k,:)/2+(1+sqrt(-1))/2;
dF(k,:)=(1-sqrt(-1))*dS(k,:)/2+(1+sqrt(-1))/2;
% Square root of the Brownian motion
dX(k,:)=(mu0+abs(W(k,:))/(2*mu0)+(-1)*dt/(8*mu0^3)).*dF1(k,:);

%% Euler-Maruyama integration of the SDEs
for i=2:N
    Winc = dW(i);
    Finc(i) = dF(k,i);
    w(k,i) = w(k,i-1)+sqrt(Winc);   % Integrate square root (SR) SDE
    z(k,i) = z(k,i-1)+Winc;         % Integrate SDE of the Brownian motion (BM)
    % Farina-Frasca-Sedehi SDE (FFS)
    v(k,i) = v(k,i-1)+(mu0+abs(Winc)/(2*mu0)+(-1)*dt/(8*mu0^3))*Finc(i);
end

wm(k,:) = abs(w(k,:));             % Module of the extracted square root
wa(k,:) = angle(w(k,:))*180/pi;    % Phase of the extracted square root in °

%% Check the correctness of the solutions of the two SDEs
y(k,1)=W(k,1);
s(k,1)=W(k,1);
m(k,1)=W(k,1);
for i=2:N
    y(k,i)=z(k,i)-z(k,i-1);                     % Jumps obtained from (BM) SDE
    s(k,i)=abs(w(k,i)-w(k,i-1))^2*sign(y(k,i)); % Jumps obtained from (SR) SDE
    % Jumps obtained from Frasca-Farina-Sedehi (FFS) SDE
    %%%%%%%%%%%%%%%%%%%%%%%%%%%%%%%%%%%%% 
    % after subtraction of the mu0 term %
    %%%%%%%%%%%%%%%%%%%%%%%%%%%%%%%%%%%%%
    m(k,i)=abs(v(k,i)-v(k,i-1))^2*sign(y(k,i))-mu0^2*sign(y(k,i));  
end

end  % End of the loop for statistical runs

%% Output

kk1 = mean(y,1);               % Mean of the solutions of SDE (BM)
kk2 = mean(s,1);               % Mean of the square of solutions 
                               % of SDE (SR)
kk0 = mean(m,1);               % Mean of the square of solutions
                               % of SDE (FFS)

figure
plot(t(1:end-1),kk1,'b*')
hold on
plot(t(1:end-1),kk2,'ro')
legend('BM','SR')

hh = mean(W,1);                          % Mean over all Brownian motions
% Frasca-Farina-Sedehi formula
ww = mean(abs(dX).^2.*dS1-mu0^2*dS1,1);  % Mean over all square roots 
                                         % of the Brownian motions
kk3 = cumsum(kk2);                       % Mean path for the SDE solutions (SR)
kk4 = cumsum(kk0);                       % Mean path for the SDE (FFS)
                               
figure
subplot(1,4,1)                 % Mean Brownian motion
plot(t(1:end-1),hh,'r')
ylim([min(hh) max(hh)])
xlabel('a)')
subplot(1,4,2)                 % Mean of the square of the square root
plot(t(1:end-1),ww,'b')
ylim([min(ww) max(ww)])
xlabel('b)')
subplot(1,4,3)                 % Mean of the squared solutions of SDE (SR)
plot(t(1:end-1),kk3,'m')
ylim([min(kk3) max(kk3)])
xlabel('c)')
subplot(1,4,4)                 % Mean of the squared solutions of SDE (FFS)
plot(t(1:end-1),kk4,'k')
ylim([min(kk4) max(kk4)])
xlabel('d)')
\end{lstlisting}

\begin{figure}[H]
  \includegraphics{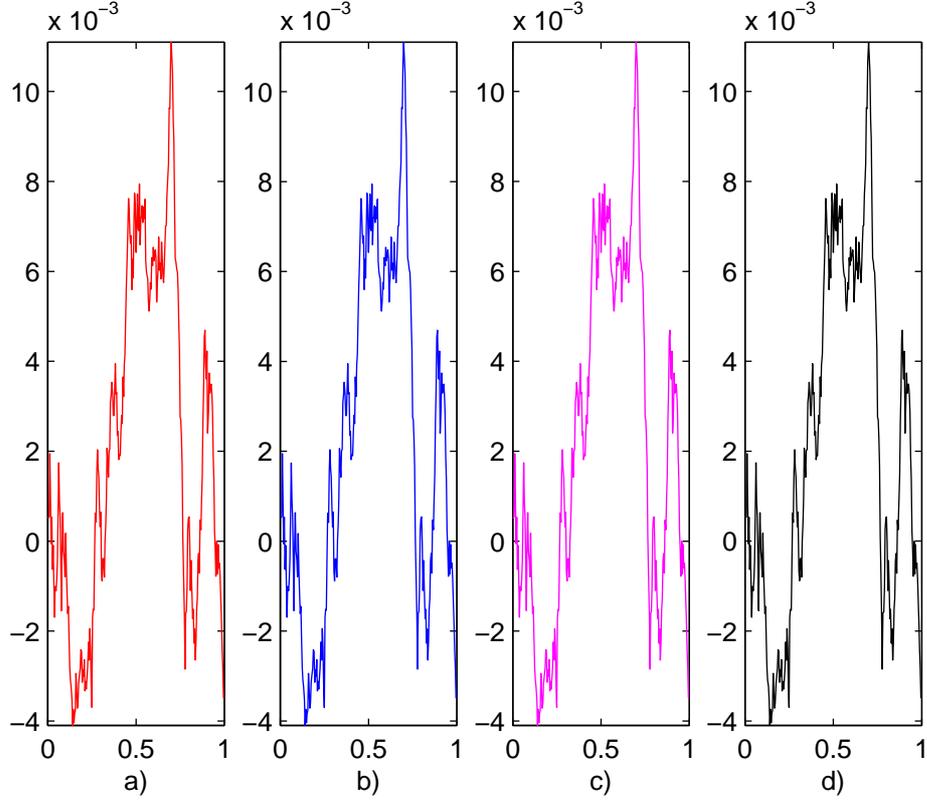}
  \caption{Monte Carlo simulation with 10000 paths for $\alpha=1/2$. a) Mean path of the Brownian motion. b) Mean square of eq.(\ref{eq:sroot}). c) Cumulative sum of the mean of the jumps from the SDE (\ref{eq:srnum}) with $\alpha=1/2$. d) Cumulative sum of the mean of the jumps from the SDE eq.~(\ref{eq:emsroot}).\label{fig:ff1}}
\end{figure}

The results are given in Fig.~\ref{fig:ff1}.
It is immediately seen that the mean path is recovered for eq.~(\ref{eq:sroot}) from the numerical solution of eq.~(\ref{eq:srnum}) and eq.~(\ref{eq:emsroot}). 
This result is really striking proving unequivocally that the square root process indeed exist and that eq.~(\ref{eq:emsroot}) as is the one proposed in \cite{far2,fra1}, eq.~(\ref{eq:sroot}), does hold. The interpretations to these formulas yielded in Sec.~\ref{sec2} are the proper ones. To achieve the exact comparison we removed the term $\mu_0^2\sgn(dW)$ (cfr. eq.(\ref{eq:Xsq})) as it can be seen from the listing of the computer code.

\begin{figure}[H]
  \includegraphics{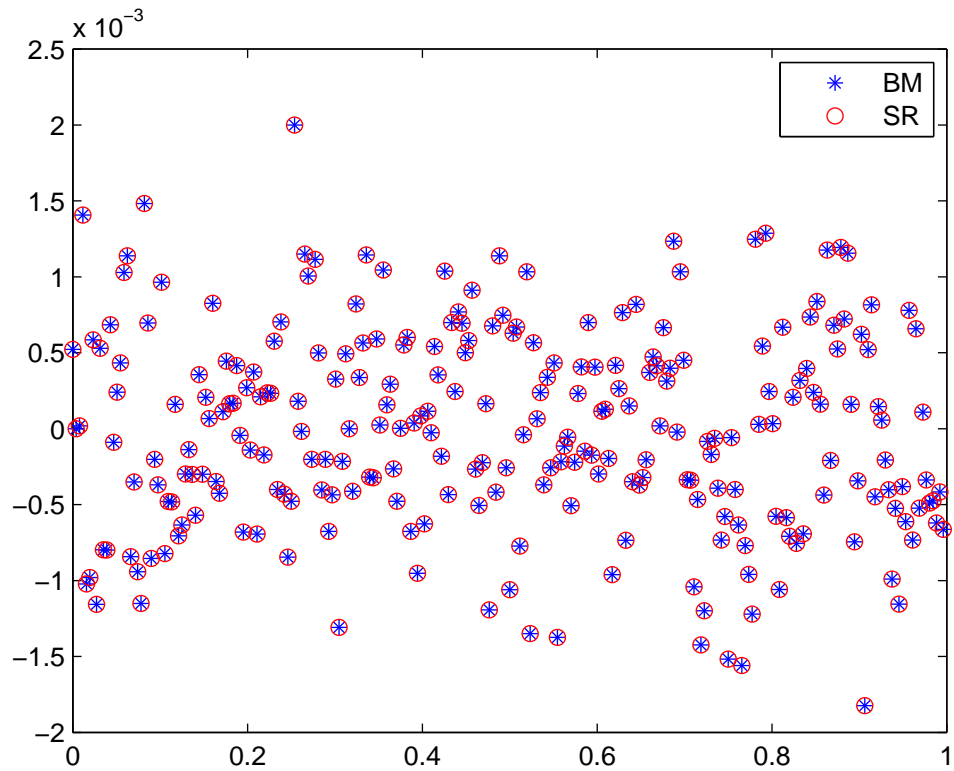}
  \caption{Monte Carlo simulation with 10000 paths for $\alpha=1/2$. Complete coincidence of the jumps of the mean Brownian path and the mean of the squares of the jumps of its square root.\label{fig:ff2}}
\end{figure}

It is also verified that the algorithm \ref{alg:BBP}, proposed in the preceding section to recover the mean Brownian path from the numerical solution of eq.~(\ref{eq:srnum}), is completely fulfilled as shown in Fig.~\ref{fig:ff2}. 

\begin{figure}[H]
  \includegraphics{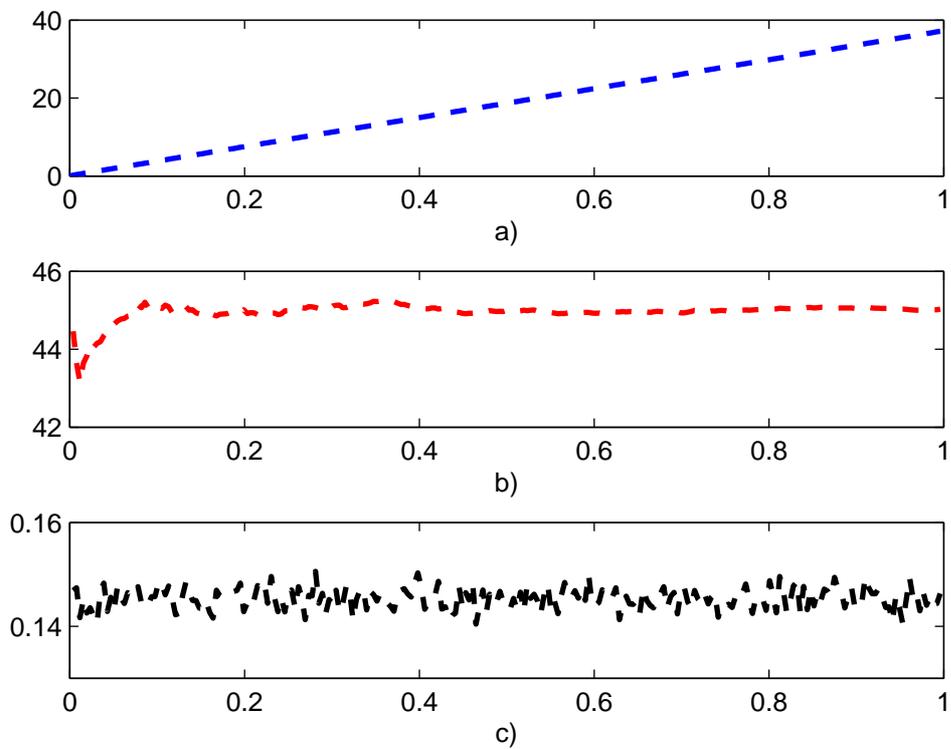}
  \caption{Components of the square root process. a) Modulus. b) Phase in degrees. c) Absolute values of the jumps.\label{fig:ff3}}
\end{figure}

It is interesting to note that the single components of the square root process have the module growing linearly with time and the inherent noisy behavior can be seen in the phase and recovered computing the jumps. These jumps describe accurately the original Brownian motion after a cumulative sum is performed. This is exemplified in Fig.~\ref{fig:ff3}.

\begin{figure}[H]
  \includegraphics{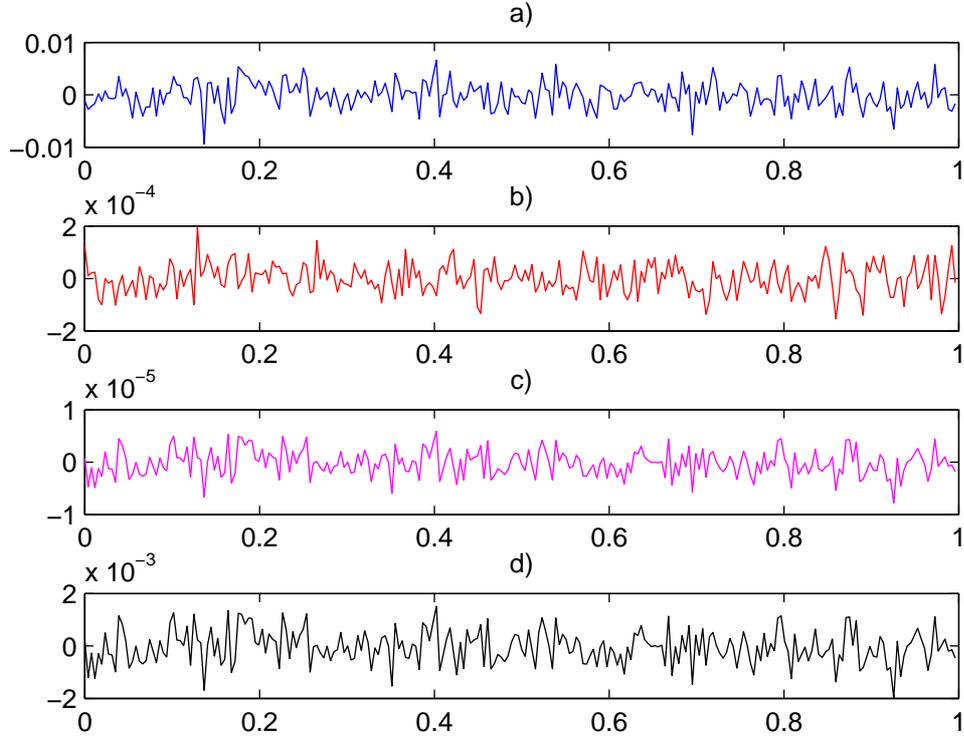}
  \caption{Smallness of contributions to the square of eq.(\ref{eq:emsroot}). a) $\sgn(W_i-W_{i-1})$. b) $(W_i-W_{i-1})^2-dt$. c) $(W_i-W_{i-1})\cdot dt$. d) $W_i-W_{i-1}$. \label{fig:ff4}}
\end{figure}

We conclude this section with a check of eq.~(\ref{eq:emsroot}) for the magnitude of $(W_i-W_{i-1})^2-dt$, $(W_i-W_{i-1})\cdot dt$ and $\sgn(W_i-W_{i-1})$. We performed a Monte Carlo simulation with 10000 paths and $dt=2^{-8}$. The results are displayed in Fig.~\ref{fig:ff4}.
This is consistent with our proof of eq.~(\ref{eq:emsroot}) that is shown to yield an excellent representation of the square root of a standard Brownian motion. This conclusion is enforced by the fact that we have chosen $\mu_0=30\gg 1$ in our numerical computations.

\section{Square root process and Clifford algebra}
\label{sec4}

Eq.(\ref{eq:sroot}) has the shortcoming that, when squared, does not yield exactly the Wiener process but adds a sign process that must be removed to recover the standard Brownian motion we started from. This problem can be solved using a Clifford algebra\footnote{A Clifford algebra is a unital associative algebra that contains and is generated by a vector space $V$ over a field $K$, where $V$ is equipped with a quadratic form $Q$. $C\ell(V,Q)$ has the condition $v^2 = Q(v)1$for all $v\in V$.} as that of Pauli matrices $C\ell_3(\mathbb{C})$ \cite{dorl}. In this case we will have for $\sigma_i\in C\ell_3(\mathbb{C})$ and $i,\ k=1,\,2,\ 3$
\begin{eqnarray}
\label{eq:cliff}
    \sigma_i^2&=&I \nonumber \\
		\sigma_i\sigma_k&=&-\sigma_k\sigma_i\qquad i\ne k,
\end{eqnarray}
these are also known as Pauli or spin matrices, and one can write
\begin{equation}
\label{eq:pE}
    dE =\sigma_i\left(\mu_0+\frac{1}{2\mu_0}|dW|-\frac{1}{8\mu_0^3}dt\right)\cdot\Phi_\frac{1}{2}+i\sigma_k\mu_0\Phi_\frac{1}{2}\qquad i\ne k
\end{equation}
being now the stochastic process $E$ an element of the given Clifford algebra. Now, using eq.(\ref{eq:cliff}), it is straightforward to obtain $(dE)^2=I\cdot dW$ and the sign process is removed. This can be seen immediately by noting that
\begin{eqnarray}
    (dE)^2&=&\sigma_i^2(\mu_0^2\sgn(dW)+dW)
		+i(\sigma_i\sigma_k+\sigma_k\sigma_i)\left(\mu_0+\frac{1}{2\mu_0}|dW|
		-\frac{1}{8\mu_0^3}dt\right)\sgn(dW)\nonumber \\
		&&-\sigma_k^2\mu_0^2\sgn(dW)
\end{eqnarray}
where use has been made of eq.(\ref{eq:Xsq}). But by eq.(\ref{eq:cliff}) is $\sigma_i^2=\sigma_k^2=I$ and $\sigma_i\sigma_k+\sigma_k\sigma_i=0$ and the result is obtained. We note anyway that a dependence from an arbitrary constant in eq.(\ref{eq:pE}) is retained. A generalization to higher Clifford algebras is immediate and an argument for future work.

This result appears really striking in view of the motivations to introduce the square root process in \cite{far2} to recover the Schr\"odinger equation. We have seen that spin Pauli matrices are needed to exactly recover the standard Brownian motion and this means that this could be the proper way to extend these ideas to recover also the Dirac equation and higher generalizations. 

\section{Conclusions}
\label{sec5}

We have shown that, using numerical integration of SDE, it is possible to generate $\alpha$--root Brownian motion. The technique of numerical integration becomes itself a definition for this kind of processes. Their existence is so proven using a numerical simulation. For the square root process, it is seen that the real and imaginary components increase without bound through their modulus but the random jumps they are made of, reproduce accurately the Brownian motion we started with. In this sense we get a significant meaning to the square root of a Brownian motion.

We provided a full accomplished Monte Carlo simulation that clearly shows that both the idea of a fractional Brownian motion and the corresponding formula in terms of the product of a Brownian motion and a Bernoulli process are sound and can be properly defined into a mathematical framework. The coincidence between numerical results we obtained is a striking evidence of this fact.

This approach seems clearly to require the introduction of a Clifford algebra to exactly recover the standard Brownian motion and this seems to support in a strong way the idea firstly put forward in \cite{far2} of a deep connection between the square root process we proved here to exist and quantum mechanics. Further studies are needed to support this idea and to see eventually how Dirac equation could emerge. 

We are developing applications in the area of signal processing for the square root process but also other fractional Wiener processes can be worth studying for this.

The consequence of this analysis is that we can extend the concepts of stochastic integral and stochastic process to include such behaviors.



\section*{References}

\begin{thebibliography}{99}
\bibitem{far2} A.~Farina, M.~Frasca, M.~Sedehi, ``Solving Schrödinger equation via Tartaglia--Pascal triangle: a possible link between stochastic processing and quantum mechanics'', Signal, Image and Video Processing, Volume 8, Number 1, 27--37 (2014).
\bibitem{fra1} M.~Frasca, ``Quantum mechanics is the square root of a stochastic process'', arXiv:1201.5091 [math-ph] (2012).
\bibitem{hard} G.~H.~Hardy, ``Divergent Series'', (AMS Chelsea Publishing, New York, 1991).
\bibitem{high} D.~J.~Higham, ``An Algorithmic Introduction to Numerical Simulation of Stochastic Differential Equations'', SIAM Review 43, 525--546 (2001).
\bibitem{okse} B.~K.~\O ksendal, ``Stochastic Differential Equations: An Introduction with Applications'', (Springer, Berlin, 2003).
\bibitem{dorl} C.~Doran, A.~Lasenby, ``Geometric Algebra for Physicists'', (Cambridge University Press, Cambridge, 2003).
\end{thebibliography}
\end{document}